\documentclass{aa}
\usepackage{graphicx}
\usepackage{natbib}

\usepackage{txfonts,textcomp}
\usepackage{color}
\usepackage{cleveref}

\newcommand{\kms}{\mbox{$\mbox{km\,s}^{-1}$}\,}

\begin{document}

\title{GIARPS simultaneous optical and infrared high-spectral-resolution observations of Cepheids}
\subtitle{I. Limb darkening from cross-correlated radial velocities}
\titlerunning{GIARPS simultaneous optical and infrared high-spectral-resolution observations of Cepheids}
\authorrunning{Nardetto et al.}

\author{N.~Nardetto \inst{1} 
\and R. S. Rathour\inst{1} 
\and M. C.~Bailleul \inst{1} 
\and V.~Hocd\'e \inst{1}  
\and P.~Kervella \inst{2,7} 
\and G. Bras \inst{2}  
\and W.~Gieren\inst{3}  
\and J.~Storm \inst{4}  
\and G.~Pietrzy\'nski \inst{5}  
\and B. Zgirski \inst{3}
\and P. Wielg\'orski \inst{5}  
\and A.~M\'erand \inst{6} 
\and A.~Gallenne\inst{8}  
\and E.~Poretti \inst{9,10}
\and M.~Rainer \inst{9} 
\and L.~ Breuval \inst{2,11} 
\and M. Duparc \inst{1}  
\and B. Apostolova\inst{1} 
\and W. Kiviaho \inst{2,7}
\and K. Sivkova \inst{2,7}  
}
\institute{Universit\'e C\^ote d'Azur, OCA, CNRS, Lagrange, France,  Nicolas.Nardetto@oca.eu   
\and LIRA, Observatoire de Paris, Universit\'e PSL, Sorbonne Universit\'e, Universit\'e Paris Cit\'e, CY Cergy Paris Universit\'e, CNRS, 92190 Meudon, France
\and Universidad de Concepci\'on, Departamento de Astronom\'ia, Casilla 160-C, Concepci\'on, Chile
\and Leibniz Institute for Astrophysics, An der Sternwarte 16, 14482, Potsdam, Germany 
\and Nicolaus Copernicus Astronomical Center, Polish Academy of Sciences, ul. Bartycka 18, PL-00-716 Warszawa, Poland
\and European Southern Observatory, Alonso de C\'ordova 3107, Casilla 19001, Santiago 19, Chile 
\and French-Chilean Laboratory for Astronomy, IRL 3386, CNRS and U. de Chile, Casilla 36-D, Santiago, Chile
\and Instituto de Alta Investigaci\'on, Universidad de Tarapac\'on, Casilla 7D, Arica, Chile
\and  INAF -- Osservatorio Astronomico di Brera, Via E. Bianchi 46, 23807 Merate (LC), Italy 
\and Fundaci\'on Galileo Galilei-INAF, Rambla Jos\'e Ana Fernandez P\'erez 7, 38712 Bre\~na Baja, TF, Spain
\and European Space Agency (ESA), ESA Office, Space Telescope Science Institute, 3700 San Martin Drive, Baltimore, MD 21218, USA
}

\date{Received ... ; accepted ...}

\abstract 
{The Baade-Wesselink method, while theoretically applicable for calibrating the distance scale via Cepheids, is not currently employed in the cosmic distance ladder. There are various versions of this method based mainly on interferometry and/or surface-brightness color relations. In all these approaches, the projection factor  used to convert the radial into the pulsation velocity remains the limiting quantity.}
{By comparing for the first time simultaneous cross-correlated optical and H-band radial velocities of five Cepheids, we aim to better understand the geometric part of the projection factor that is directly linked to the limb darkening of the stars. The five Cepheids were observed  with the  HARPS-N and GIANO-B instruments mounted at the Telescopio Nazionale Galileo in the GIARPS configuration.} 
{A two-parameter linear fit was applied to the simultaneous radial velocity measurements of each star in order to extract the amplitude ratio of both curves ($f_\mathrm{VH}$) as well as the difference in mean velocities. 
}  
{We find a mean value of  $f_\mathrm{VH}$ of 1.056 with a root mean square over the five stars of 0.016, which means that the V-band radial velocity amplitudes are around 5\% larger than the ones obtained in the H band. In addition, we find that the H-band radial velocity curves have an average (i.e., a $\gamma$ velocity) of about 1 \kms blueshifted compared to the visible ones, except for one star, X Cyg, which has a blueshift of 0.4 \kms. At first order, the $f_\mathrm{VH}$ values are directly linked to the ratio of limb darkening coefficients in both bands (or equivalently the ratio of geometrical projection factors). The $f_\mathrm{VH}$ values are found to be consistent with expectations from static stellar atmosphere models. The difference in the $\gamma$ velocity is attributed to the interplay of convection and the deeper location of H-band line-forming regions.} 
{
This work shows that the differential limb darkening of Cepheids in visible and H bands, and consequently the geometrical part of the projection factor, follows what is expected from stellar atmosphere models. The current dispersion of about 8\% in the projection factors (for a given pulsation period) is most likely due to poorly understood dynamical effects in the atmosphere of Cepheids.}
\maketitle
\nolinenumbers

\section{Introduction}\label{s_Introduction}

Cepheids are iconic astrophysical objects for deriving cosmic distances and determining the Hubble-Lema\^itre constant, a quest that has driven intense research for over a century -- from early foundational work \citep{hubble29b, sandage72a, sandage72b} to modern advancements \citep{freedman01, riess22, breuval24, freedman25, ho2026}.
This is thanks to the relation between their pulsating period and their luminosity \citep{leavitt1912}. But their pulsation has another consequence: they allow one to apply the so-called parallax-of-pulsation method of distance determination, which could potentially open an alternative road to the Hubble-Lemaitre constant \citep{nardetto23b}. The fundamental principles first developed by \citet{baade26} and \citet{wesselink46} have been improved with time up to their most recent versions, based on either interferometric measurements \citep{lane00, kervella04a, merand05} or on the surface-brightness color relation (SBCR; \citealt{storm11a, storm11b}), which requires a specific interferometric calibration \citep{kervella04c, bailleul25}. A global approach that combines several photometric bands, velocimetry, and interferometry was developed ten years ago \citep[SPIPS; ][]{merand15}. The fundamental principle of the parallax-of-pulsation method is rather simple: distances are computed using measurements of the angular diameter (directly from interferometry or indirectly through SBCR) over the whole pulsation period along with the stellar radius variations deduced from the integration of the pulsation velocity, $V_{\rm puls}$. The pulsation velocity is linked to the observed radial velocity (RV) by the projection factor $p=V_{\rm puls}/RV$. There are several ways of defining the projection factors  \citep{nardetto04} and the technique used to derive the RV (pixel minimum, centroid, Gaussian fit, and cross-correlation over thousands of lines) needs to be taken into account \citep{nardetto06a, nardetto09, borgniet19}. For the first time, \citet{nardetto07} divided the projection factor into a geometrical part intensively studied in the past \citep{vanhoof52, getting34, burki82} and a dynamical part (including atmosphere velocity gradient) that usually requires, to be studied in detail, high-signal-to-noise and high-spectral-resolution observations \citep{nardetto17} or hydrodynamical simulations \citep{sabbey95, nardetto04, vasilyev17}. 
A more direct way to study the projection factor was proposed by \citet{merand05} by inversing the methodology: as the projection factor is directly linked to the distance, if the distance is known (from trigonometric parallaxes for instance), the projection can be derived. This has been done recently on a large sample of stars using the SPIPS approach \citep{gallenne17, trahin21} and a high dispersion of about 8\% has been found in the projection-factor--period-relation,  which remains difficult to explain physically. The projection factor can also be derived for different kinds of pulsating stars, which also helps to better understand their physical properties: $\beta$-Cephei stars \citep{nardetto13}, $\delta$-Scuti \citep{nardetto14}, RR Lyrae \citep{bras24, zgirski24}, and type II Cepheids \citep{wielgorski24}. 

The dynamical structure of the atmosphere of Cepheids has been well studied in the optical 
\citep{sanford56, bell64, karp75c, sasselov90,  wallerstein15, nardetto06a, anderson14, anderson16, nardetto17, hocde20b}, while only a few spectroscopic observations are available in the infrared. At 1.1~$\mu$m, \citet{sasselov89} 
used the Fourier Transform Spectrometer (FTS) at the Canada-France-Hawaii Telescope (CFHT; \citealt{maillard82}) to monitor two Cepheids, X~Sgr ($P=$7.013~d) and $\eta$~Aql ($P=$7.177~d). They reported RV 
curves with systematically larger amplitudes (20--35\%) in the infrared than in the optical domain. However, X~Sgr is now known to be an atypical Cepheid, with the possible presence of shockwaves \citep{mathias06} while their phase coverage regarding $\eta$ Aql was particularly sparse to derive the amplitude velocity at 1.1~$\mu$m with high precision.
These results were later further analyzed in terms of line asymmetry in a subsequent study based on a larger set of infrared 
spectroscopic data \citep{sasselov90}. Later, \citet{sasselov94a, sasselov94b, sasselov94c} 
studied the HeI 10830~\AA\,  spectral line to model the chromospherical structure of  
Cepheids. More recently, \citet{nardetto18b} compared the RV curve of the long-period Cepheid $\ell$ Car considering an iron line in the visible together with a calcium line in the K band. They found that the infrared and visible RV curves have rather similar amplitudes, but with a different shape at maximum receding velocity, 
while a phase shift was observed, indicating that the infrared line probably forms in a higher atmospheric region. A systematic shift of $0.5 \pm 0.3$ km/s between the infrared and visible RV curves (the infrared line being redshifted compared to the visible one) was attributed to the combination of granulation and to the fact that the calcium line was forming again in an upper part of the atmosphere compared to the visible line.

This work is part of the Unlockpfactor project, which aims to better understand the projection factor \citep{nardetto23a}. In this study we explore the H-band spectroscopic properties of Cepheids. To do this we observed five Cepheids with the HARPS-N and GIANO-B instruments combined  in the GIARPS configuration mounted at the Telescopio Nazionale Galileo \citep{claudi16}. Using this dataset, two studies have already focused on the He~I~10830 \AA\ spectral line and the effective temperature determination of Cepheids, respectively \citep{andri23, kov23}. In a third study, \citet{nardetto24a} reported the 2019 HARPS-N RV measurements of $\delta$~Cep, presented clear evidence of the presence of an inner companion, and derived the orbital parameters of the system using a Bayesian approach. In this work, we present the GIARPS dataset in Sect.~\ref{s_GIARPS}. In Sect.~\ref{s_RVccg}, 
we compare the cross-correlated RV curves obtained in the V and H bands, respectively, which then leads to a discussion about the geometric projection factor of the Cepheids in our sample in Sect.~\ref{s_po}. We conclude in Sect.~\ref{s_conclusion}.

\section{The GIARPS observations}\label{s_GIARPS}

\begin{table*}
\begin{center}
\caption{Ephemeris for the Cepheids in our sample.}\label{tab_ephemeris}
\begin{tabular}{llcccc}
\hline \hline \noalign{\smallskip}
Star & P(O-C)   &   EpP(O-C)     & $\frac{dP}{dt}$ &  $\sigma_\frac{dP}{dt}$  \\ 
\hline
Del Cep  & 5.3663671 & 2412028.2560 & -0.000105379 & 0.000003963 \\
S Sge    & 8.3820860 & 2442677.7931 & 0.000072957 & 0.000018250 \\
X Cyg    & 16.3861940 & 2448875.0560 & -0.000538121 & 0.000161831 \\
SV Vul   & 45.0068000 & 2443711.0618 & -0.279892000 & 0.001885080 \\
S Vul    & 68.0000000 & 2444138.0057 & 0.472032000 & 0.009246440 \\
\hline
  & d & d & 0.01d/yr & 0.01d/yr \\
\hline

\hline \noalign{\smallskip}
\end{tabular}
\end{center}
\textbf{Notes:} The ephemerides are from \cite{Cso22}. P(O-C) and  EpP(O-C) are the period and the reference epoch assumed for the O-C diagram, and $\frac{dP}{dt}$ is the period change rate with its corresponding uncertainty, $\sigma_\frac{dP}{dt}$. 
\end{table*}

\begin{table}
\begin{center}
\caption{Orbital parameters of $\delta$ Cep and S Sge.}\label{tab_binaries}
\begin{tabular}{lll}
\hline \hline \noalign{\smallskip}
 Parameter    & S Sge  &   $\delta$ Cep     \\ 
\hline
    P  [yr] & 1.849 & 9.318  \\
    T [d] & 2439902.3 & 2445077.810  \\ 
    e  & 0.23  & 0.711 \\
    a ["] & 1.559 & 0.029 \\ 
    $\omega$ [°]  & 203.1 & 230.4 \\ 
    i [°] &  &  123.8 \\
    $V_\mathrm{0}$ [\kms] & -10.3 & -17.28 \\
    $\frac{f}{\pi}$ [pc] & 0.14 & 27.25 \\
    K  [\kms] &  15.5.  &  1.509  \\
\hline
\hline \noalign{\smallskip}
\end{tabular}
\end{center}
\textbf{Notes:} Orbital parameters of $\delta$ Cep and S Sge from \citet{nardetto24a, nardetto24b} and \citet{evans93}, respectively. The parameters are as follows: P (period), T (epoch of periastron passage), e (eccentricity), a (semimajor axis), $\omega$ (argument of periastron), i (inclination), $V_\mathrm{0}$ (systemic velocity), $\frac{f}{\pi}$ (fractional mass of the system divided by parallax, in parsecs), and K (semi-amplitude of the RV curve).
\end{table}

\begin{figure}[htbp]
\begin{center}
\resizebox{0.75\hsize}{!}{\includegraphics[clip=true]{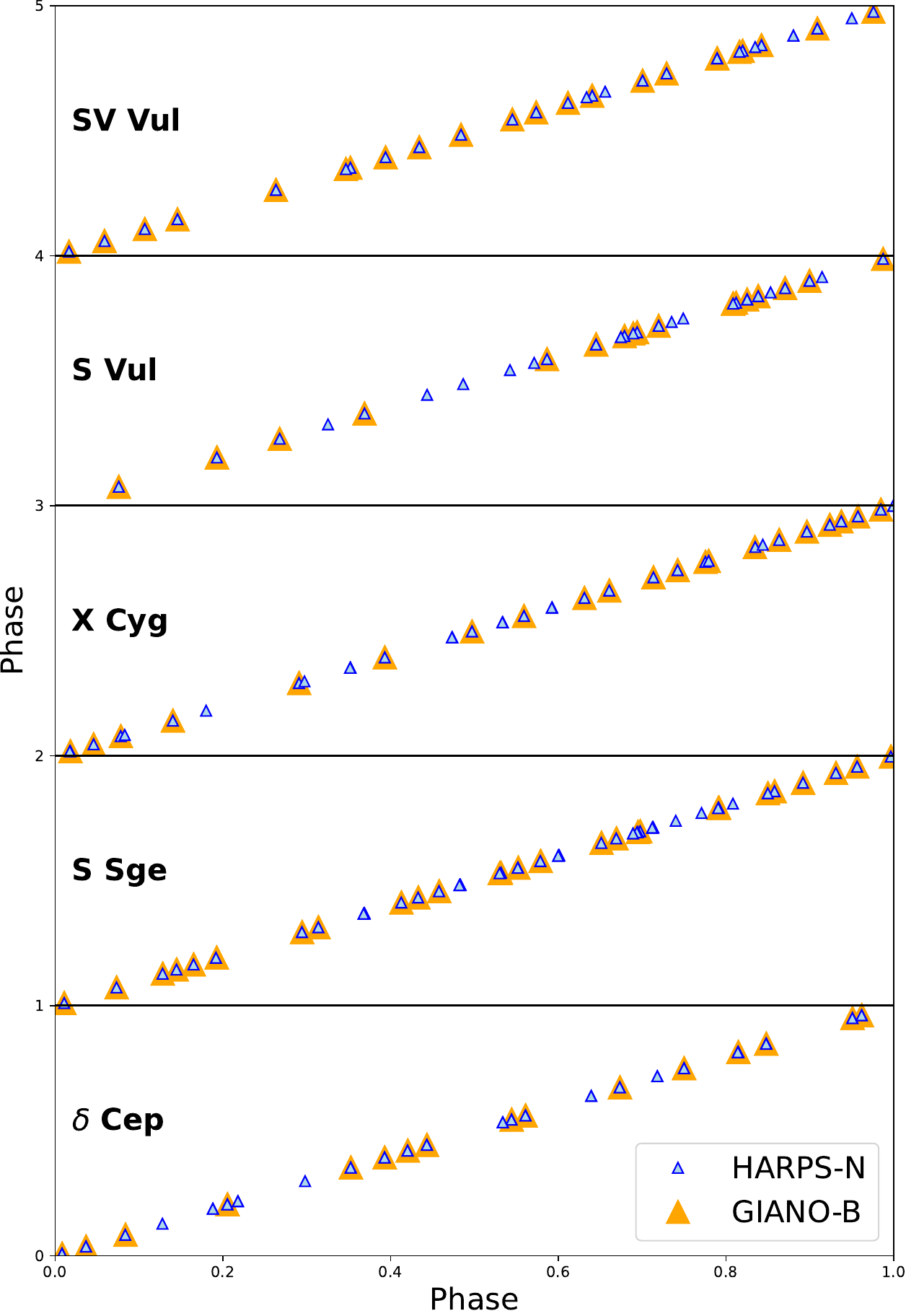}}
\end{center}
\caption{Phase coverage in the optical (HARPS-N) and in the H-band infrared (GIANO-B) are represented for the five Cepheids. For each GIANO-B observation indicated in this plot, a simultaneous HARPS-N observation has been secured, while for some HARPS-N observation, no GIANO-B data are available.}
\label{fig_prof}
\end{figure}

HARPS-N is a high-precision radial-velocity spectrograph installed at the Italian Telescopio Nazionale Galileo (TNG), a 3.58-meter telescope located at the Roque de los Muchachos Observatory on the island of La Palma, Canary Islands, Spain \citep{co12}.  HARPS-N is the northern hemisphere counterpart of the similar HARPS instrument installed at the ESO 3.6 m telescope at La Silla Observatory in Chile. The instrument covers the wavelength range from 3800 to 6900~\AA\ (V band) with a resolving power of $R \simeq 115000$. A total of 143 spectra were secured at different epochs between 27 April and 22 September 2019 in the framework of the OPTICON proposal 2019A/5. The survey includes five Cepheids: the prototype $\delta$~Cep, S~Sge,  X~Cyg, SV~Vul, and S~Vul. For 102 epochs (among 143) simultaneous spectra were obtained with the GIANO-B instrument. 
GIANO-B is a high-resolution near-infrared (NIR) echelle spectrograph. Originally designed as a cryogenic spectrograph (GIANO) for direct telescope feeding, it was later upgraded to operate in GIARPS mode, where a dichroic splits the light between the HARPS-N fibers and the GIANO-B spectrograph, allowing simultaneous observations with HARPS-N and covering a wide spectral domain from visible to NIR. Unlike its predecessor, GIANO-B is a slit spectrograph, avoiding the modal noise issues that affected the fiber-fed GIANO. GIANO-B is thus an optimized NIR echelle spectrograph that can yield, in a single exposure, 0.9-2.5 micron (H-band) spectra at R$\simeq$50,000.

The final products of the HARPS-N and GIANO-B data reduction software (DRS), installed at TNG in on-line mode, are background-subtracted, cosmic-ray-corrected, flat-fielded, and wavelength-calibrated spectra (see respectively, \citealt{co12} and \citealt{Rainer18}). Telluric lines are corrected using the MolecFit sky modeling tool \citep{smette15, kausch15}. To calculate the cross-correlated radial velocities with HARPS-N, we used the \emph{iSpec} tool with a G2V template \citep{bc14, bc19}. As shown in \citet{nardetto23a}, the choice of template (G2V or F6I) has no significant impact on the derived radial velocities. A Gaussian fit was then applied to the cross-correlation function to derive the RV ($RV_\mathrm{cc-g}$) and its uncertainty. For GIANO-B, we opted for a G2V star mask instead of a stellar atmosphere template to minimize potential systematic errors inherent in theoretical models. We verified that the results presented here are consistent within 1$\sigma$ for all stars, except for X Cyg, which shows differences of about 2$\sigma$ in terms of the ratio between V and H band RV amplitudes.

To calculate the pulsation phase of each spectrum, we used the recent ephemeris from \citet{Cso22} indicated in Tab.~\ref{tab_ephemeris}. 
For each star, we folded all measurements over a unique cycle, assuming that there are no cycle-to-cycle variations \citep{anderson14, anderson16}, which seems to be confirmed for all stars, as is discussed in next section. The pulsation phase coverage over the cycle of each Cepheid is shown in Fig.~\ref{fig_prof}. This figure also shows the phases for which simultaneous HARPS-N and GIANO-B measurements were secured. These simultaneous data are particularly precious to study the pulsating atmosphere of Cepheids. 

Two stars in our sample are binaries: $\delta$ Cep and S Sge. Regarding $\delta$ Cep, we consider the orbital parameters found in \citet{nardetto24a, nardetto24b}. For S Sge, we use the orbital parameters found by \citet{evans93}. The parameters for both stars are reported in Tab.~\ref{tab_binaries}. The resulting cross-correlation radial velocities ($RV_\mathrm{cc-g}$) together with their associated uncertainties are presented in Table A1 to A5 for the five Cepheids in our sample\footnote{Table A.1 to Table A.5  are only available in electronic form at the CDS via anonymous ftp to cdsarc.u-strasbg.fr (130.79.128.5) or via http://cdsweb.u-strasbg.fr/cgi-bin/qcat?J/A+A/.}. 

\begin{figure*}[htbp]
\centering
\includegraphics[width=0.42\textwidth]{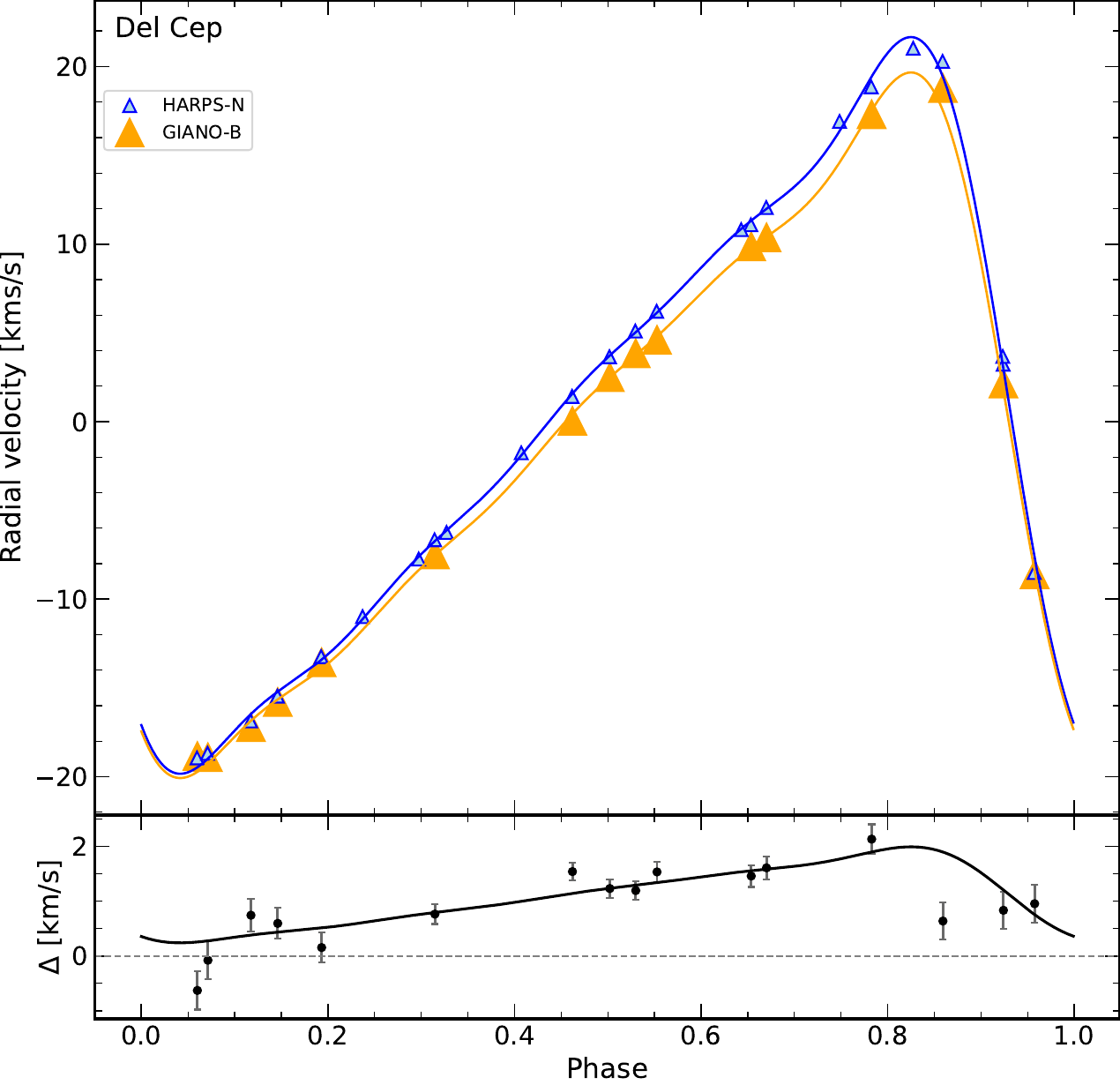}
\hspace{0.05\textwidth}
\includegraphics[width=0.42\textwidth]{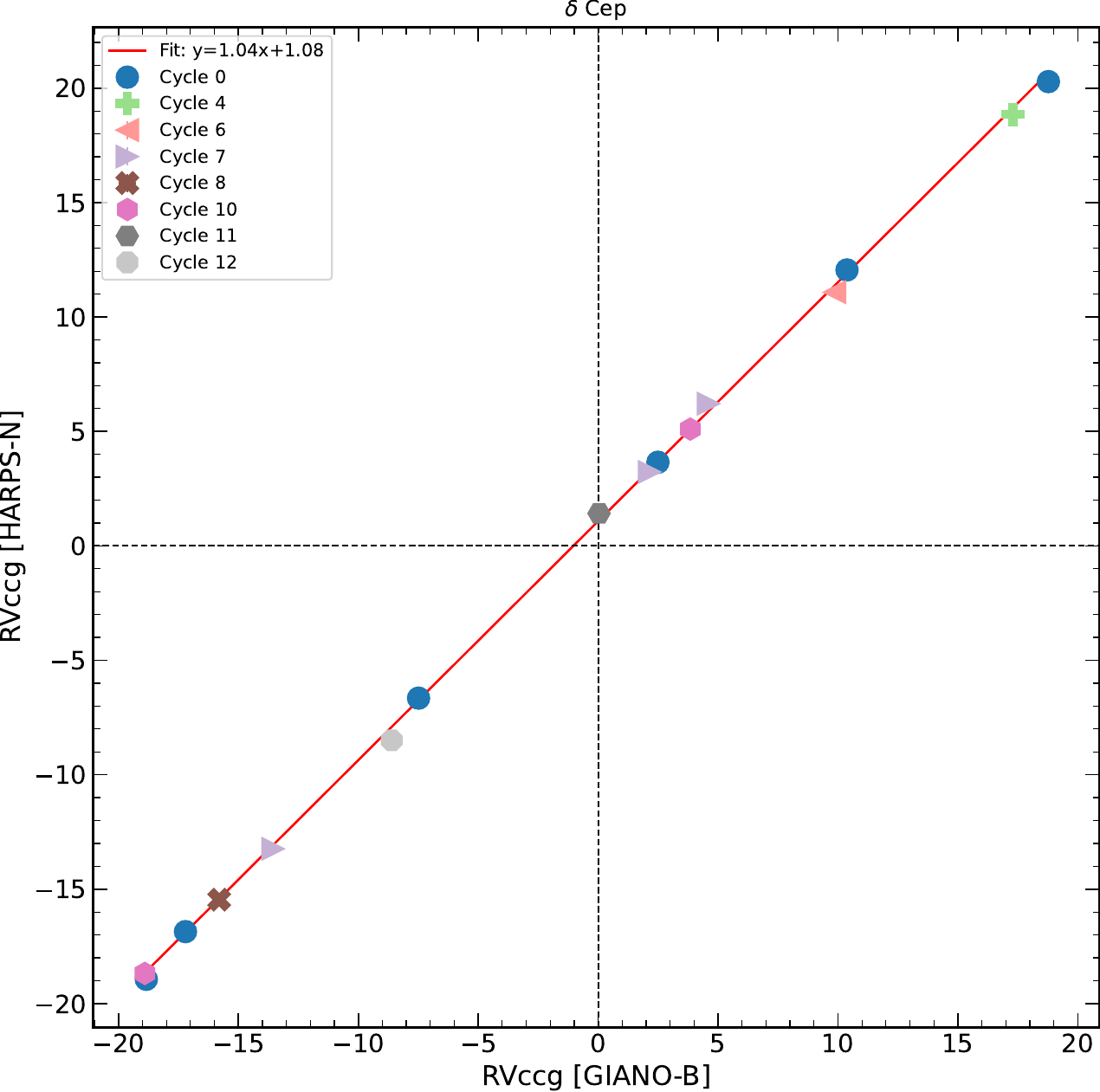} \\[1ex]
\includegraphics[width=0.42\textwidth]{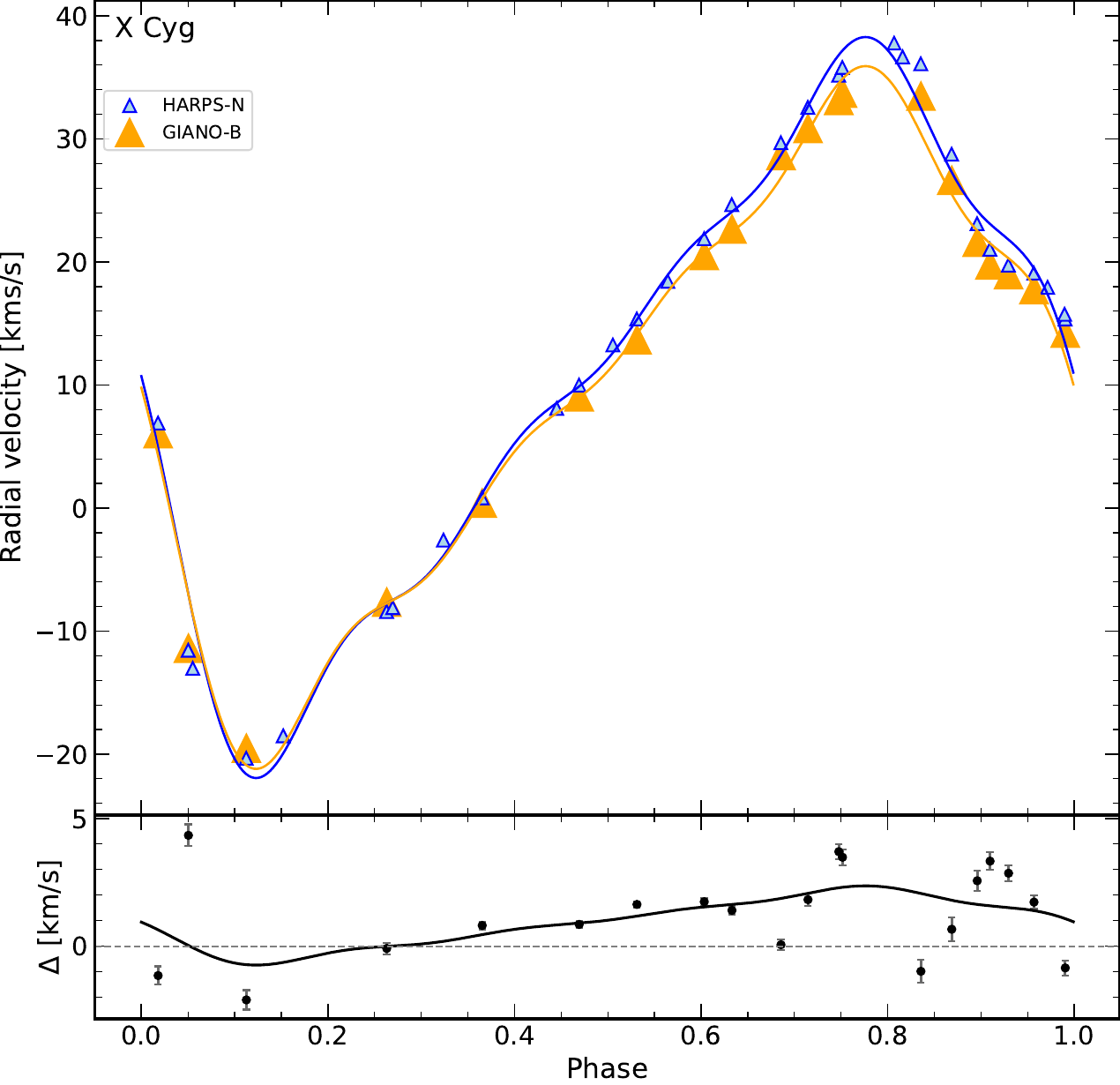}
\hspace{0.05\textwidth}
\includegraphics[width=0.42\textwidth]{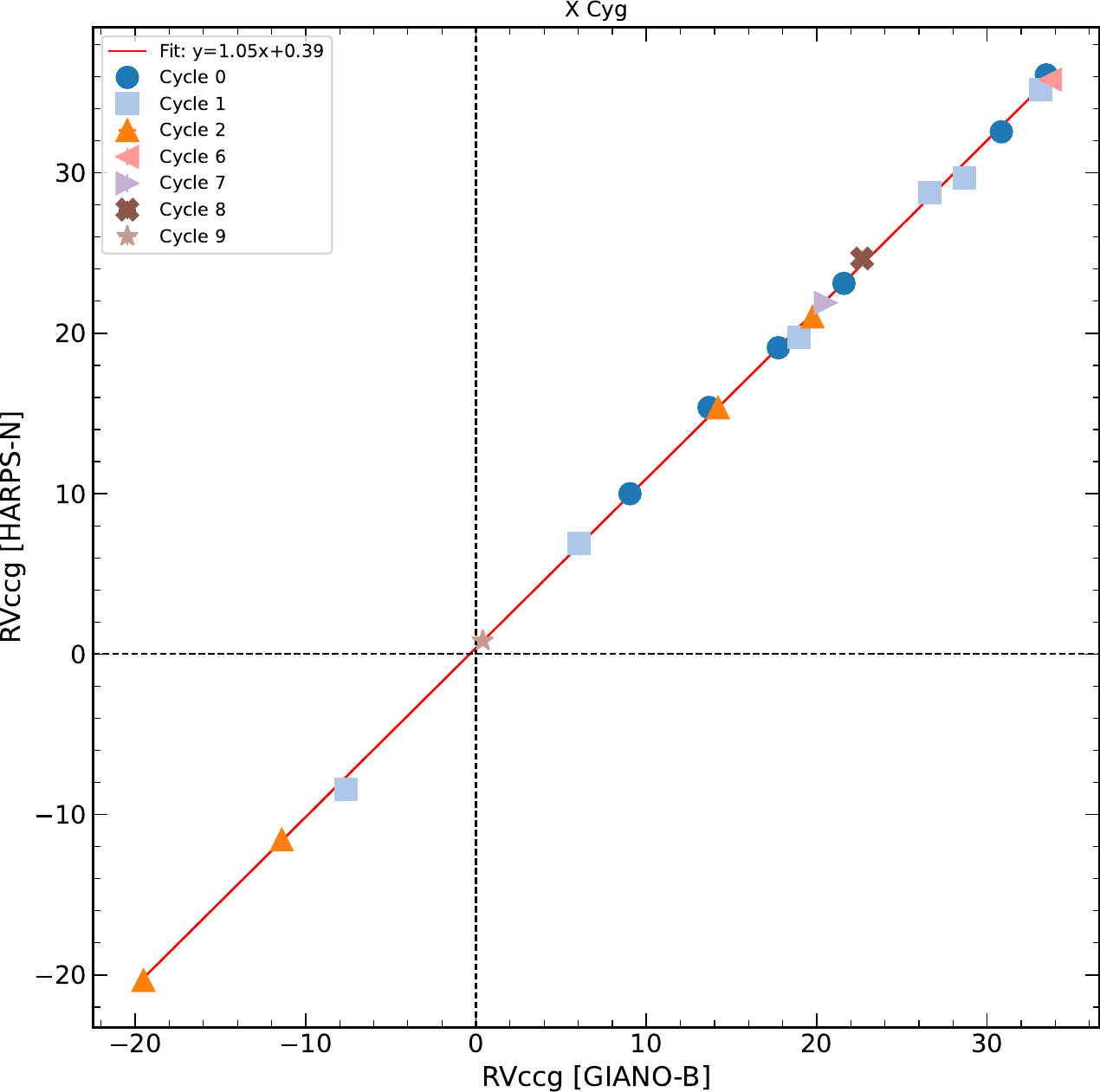} \\[1ex]
\includegraphics[width=0.42\textwidth]{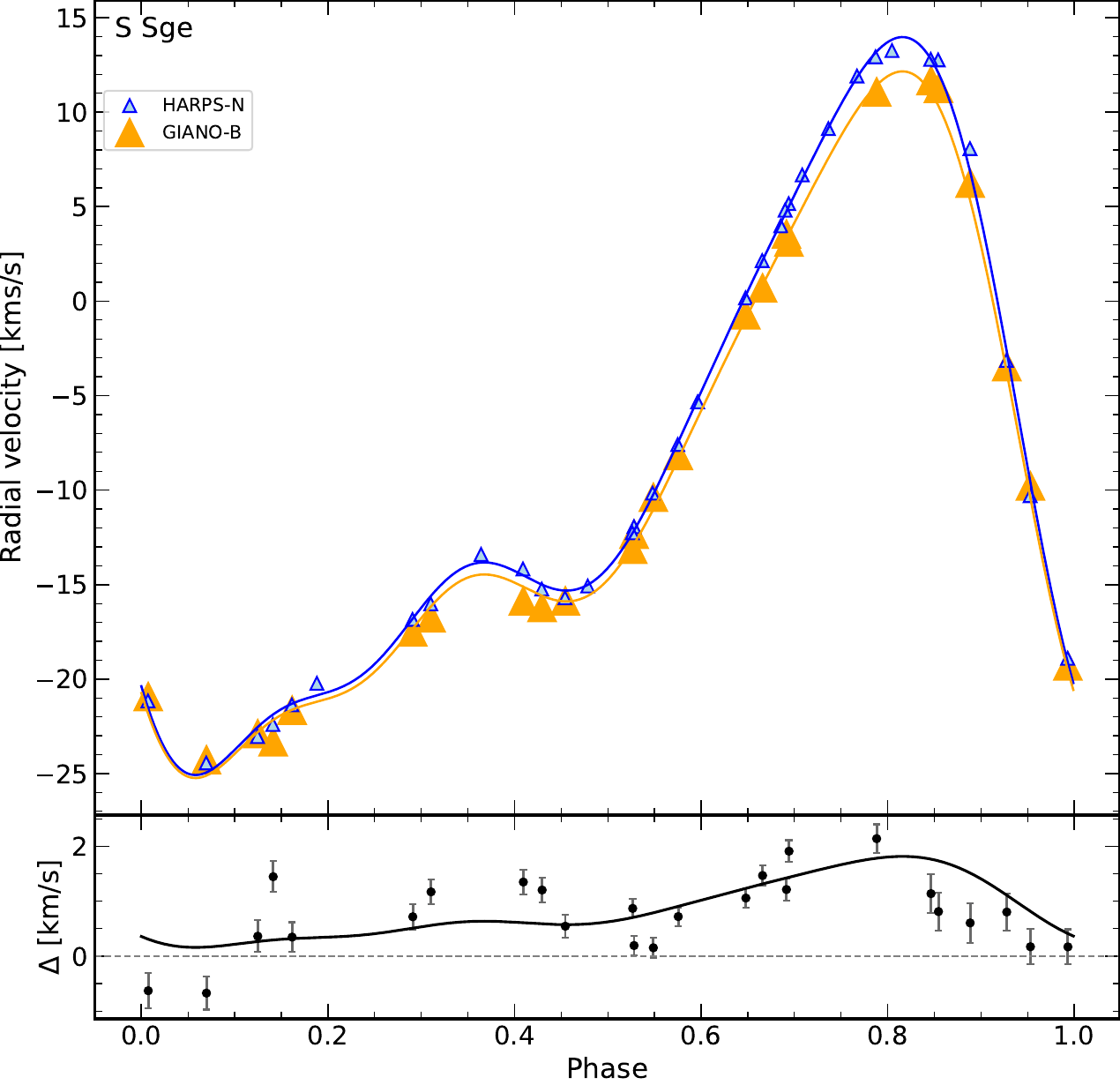}
\hspace{0.05\textwidth}
\includegraphics[width=0.42\textwidth]{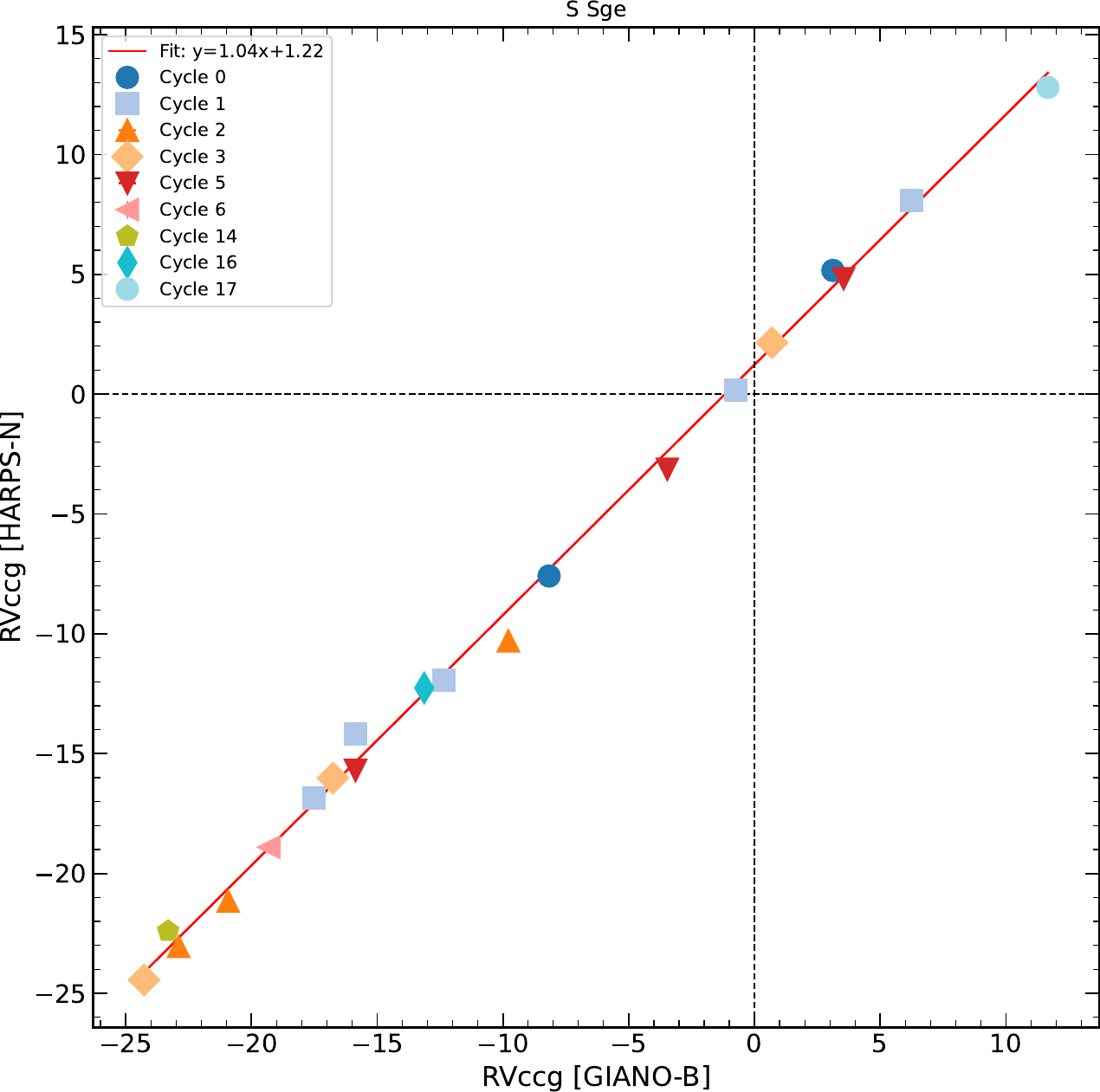}
\caption{Cross-correlated radial velocities ($RV_\mathrm{ccg}$) for $\delta$ Cep, X Cyg and S Sge. \textit{Left panels}: Comparison of HARPS-N (blue triangles) and GIANO-B (orange triangles) measurements. The sub-panels show the corresponding difference. \textit{Right panels}: $RV_\mathrm{ccg}$ from GIANO-B as a function of the ones from HARPS. The different cycles are indicated together with the best linear fit (red line).}
\label{fig_RVcc1}
\end{figure*}

\begin{figure*}[htbp]
\centering
\includegraphics[width=0.45\textwidth]{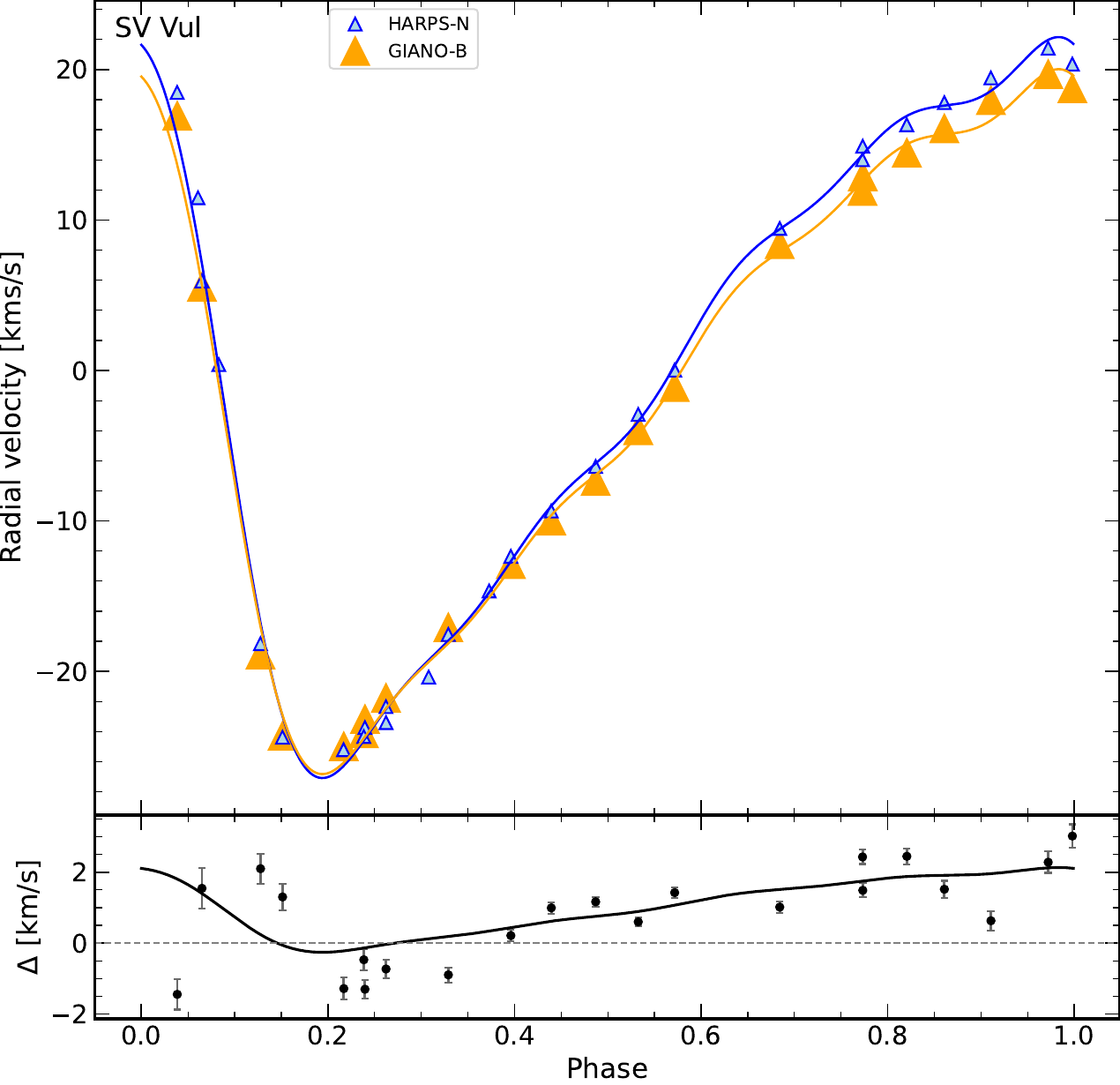}
\hspace{0.05\textwidth}
\includegraphics[width=0.45\textwidth]{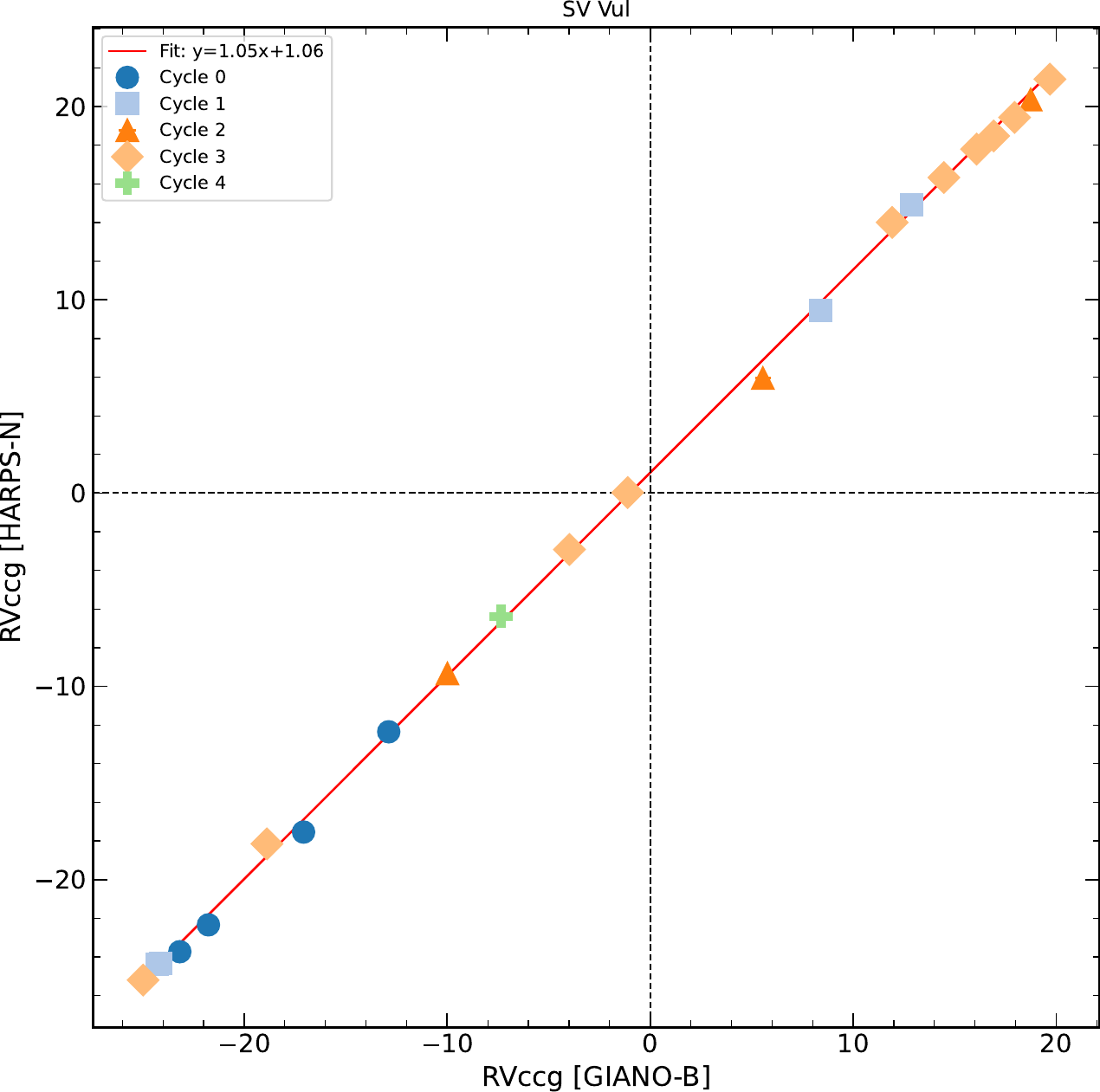}
\includegraphics[width=0.45\textwidth]{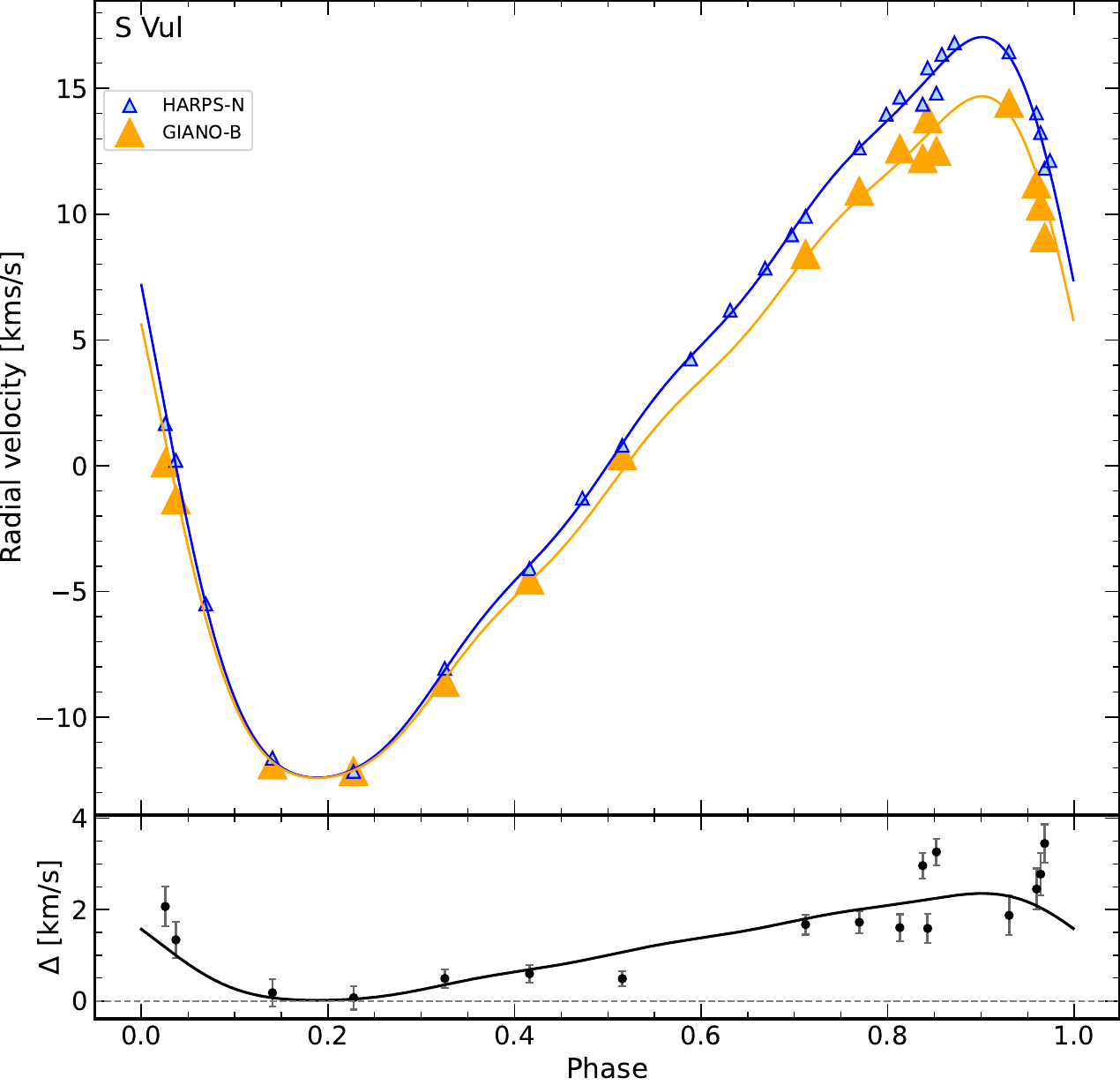}
\hspace{0.05\textwidth}
\includegraphics[width=0.45\textwidth]{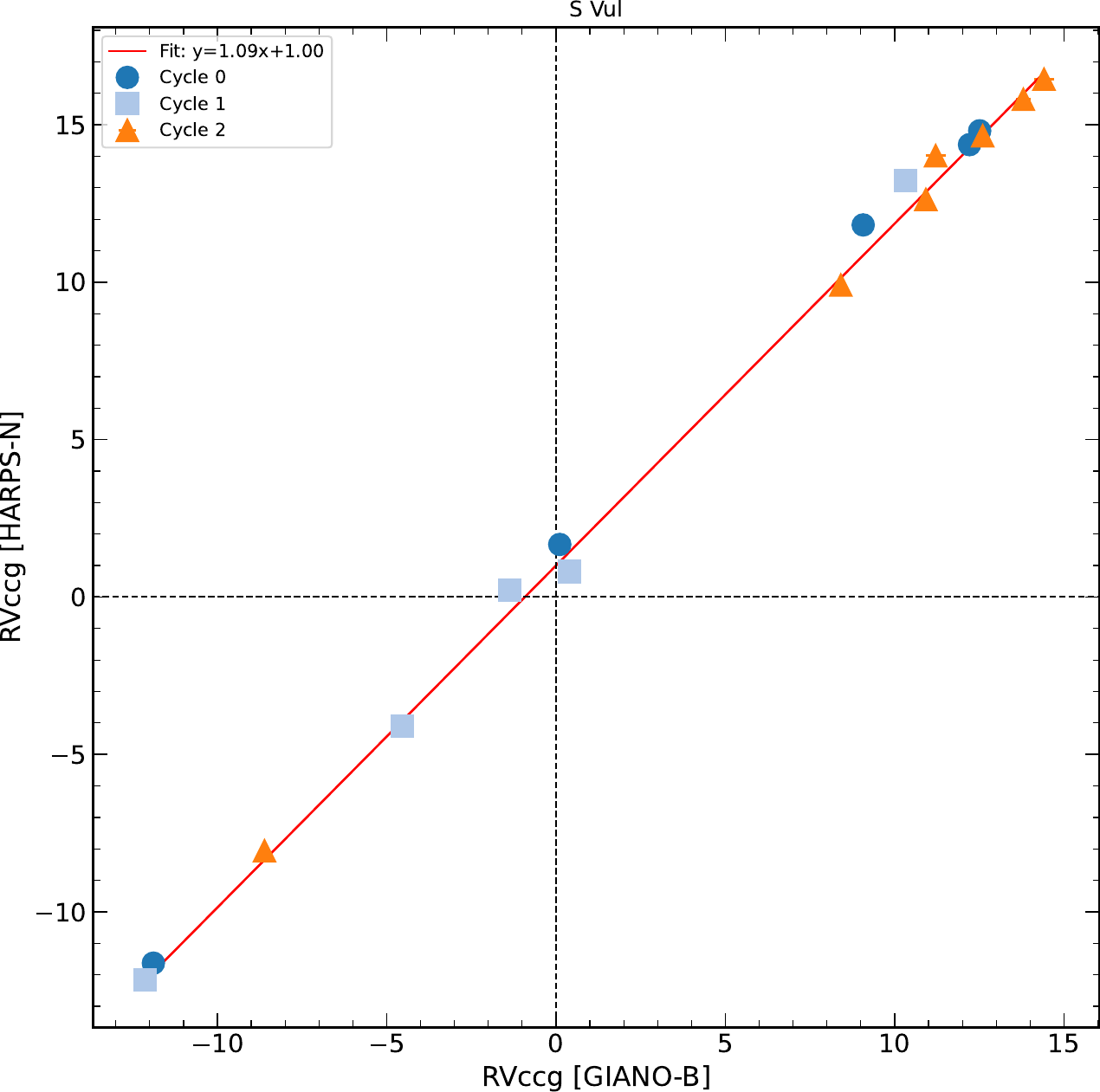} \\[1ex]
\caption{Same as Fig. \ref{fig_RVcc1} but for SV Vul and S Vul.}
\label{fig_RVcc2}
\end{figure*}

\section{Comparing the cross-correlated radial velocities}\label{s_RVccg}

In this section we compare the cross-correlation radial velocities (based on a Gaussian fit)  obtained with HARPS-N ($RV_{\mathrm{cc-g [V]}}$)  and GIANO-B ($RV_{\mathrm{cc-g [H]}}$) instruments. To do this comparison, we focus on simultaneous measurements only. We applied a two-parameter orthogonal distance regression (ODR) algorithm between $RV_{\mathrm{cc-g [V]}}$ and $RV_{\mathrm{cc-g [H]}}$ using

\begin{equation}\label{Eq_ratio}
RV_{\mathrm{cc-g [V]}}  =  f_\mathrm{VH} * (RV_{\mathrm{cc-g [H]}}) + \Delta V_{\mathrm{\gamma}}
,\end{equation}

\noindent where  $f_\mathrm{VH}$ is the ratio of the two velocity measurements, while $\Delta V_{\mathrm{\gamma}}$ is the $\gamma$-velocity offset between the V and H bands. The left panels of Fig. \ref{fig_RVcc1} and \ref{fig_RVcc2} show  the $RV_{\mathrm{cc-g [V]}} $ (open blue triangles) and $RV_{\mathrm{cc-g [H]}}$ (orange triangle) quantities as a function of the pulsation phase, together with the corresponding differences (lower sub-panels). In these sub-panels, the uncertainties correspond to the sum of the uncertainties associated with visible and H-band radial velocities. On the right panels of the same figures, we plot for each star $RV_{\mathrm{cc-g [H]}}$ as a function of $RV_{\mathrm{cc-g [V]}}$ together with the corresponding pulsating cycle of observations in order to investigate possible cycle-to-cycle variations. The red line on these plots corresponds to the ODR fit. The results of the fit are listed in Tab~\ref{tab_res} together with the reduced $\chi^2$ and the degeneracy between the two parameters. On the left panels of Figs. \ref{fig_RVcc1} and \ref{fig_RVcc2}, we interpolate the RV curve in the V band using a five-parameter Fourier fit (solid blue line), while the orange curve is a rescale and shift of the blue curve following the parameters of the fit ($f_\mathrm{VH}$ and $\Delta V_{\mathrm{\gamma}}$). These interpolations are indicative as they are not used in the following. In the sub-panels, the black lines correspond to the difference between the solid blue and orange lines.

To compare the RV curves, it would have been possible to proceed slightly differently. Indeed, we could have removed first their respective $\gamma$ velocities to each curve before calculating the ratio. However, such an approach would have been particularly sensitive to the interpolation procedure in the visible and H bands, respectively. Instead, we prefer to take the benefit of simultaneous data without adding any potential systematical uncertainty due to interpolation by applying the two-parameter fit of Eq.~\ref{Eq_ratio}, with the only drawback that the two-parameter might be slightly correlated as shown in Tab~\ref{tab_res}. Nevertheless, the results obtained with both methods provide consistent 1$\sigma$ $f_\mathrm{VH}$ values, and therefore the choice of the methodology does not change the conclusions of this study.

\begin{figure}[htbp]
\resizebox{1.0\hsize}{!}{\includegraphics[clip=true]{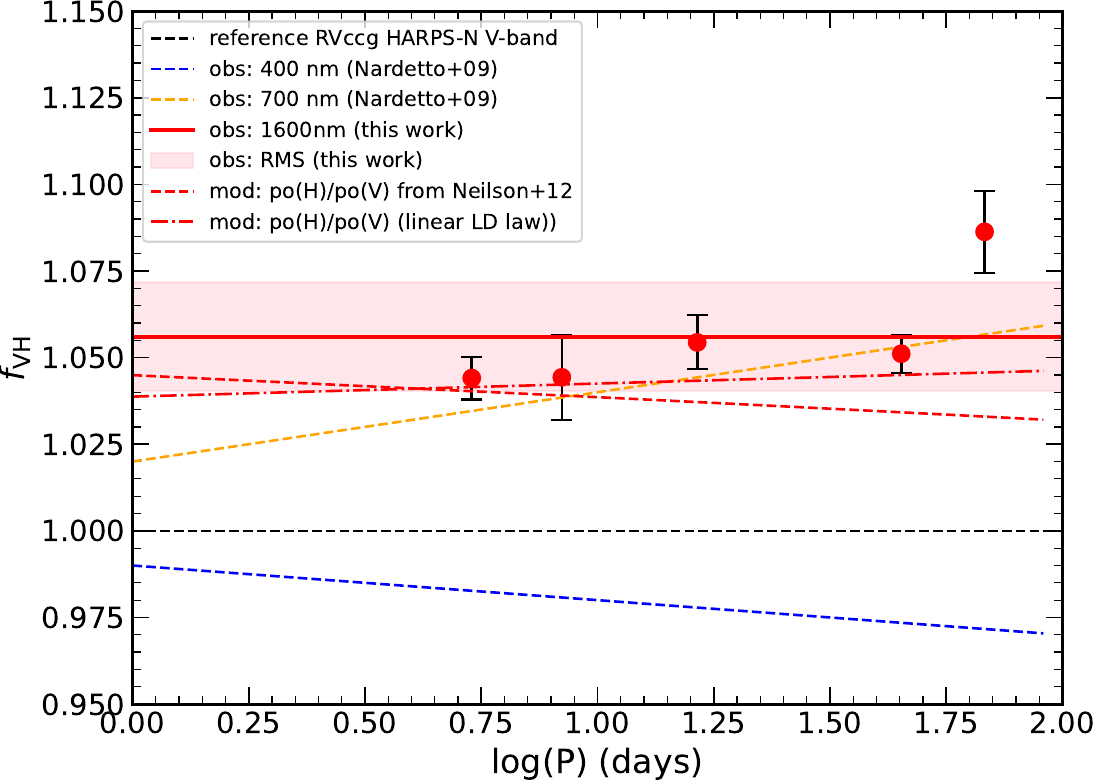}}
\caption{Amplitude ratio of the cross-correlated radial velocities in V and H band as a function of the logarithm of the period (red dots) together with some previous results and models (see text for explanations).}
\label{fig_fc}
\end{figure}

\begin{figure}[htbp]
\resizebox{1.0\hsize}{!}{\includegraphics[clip=true]{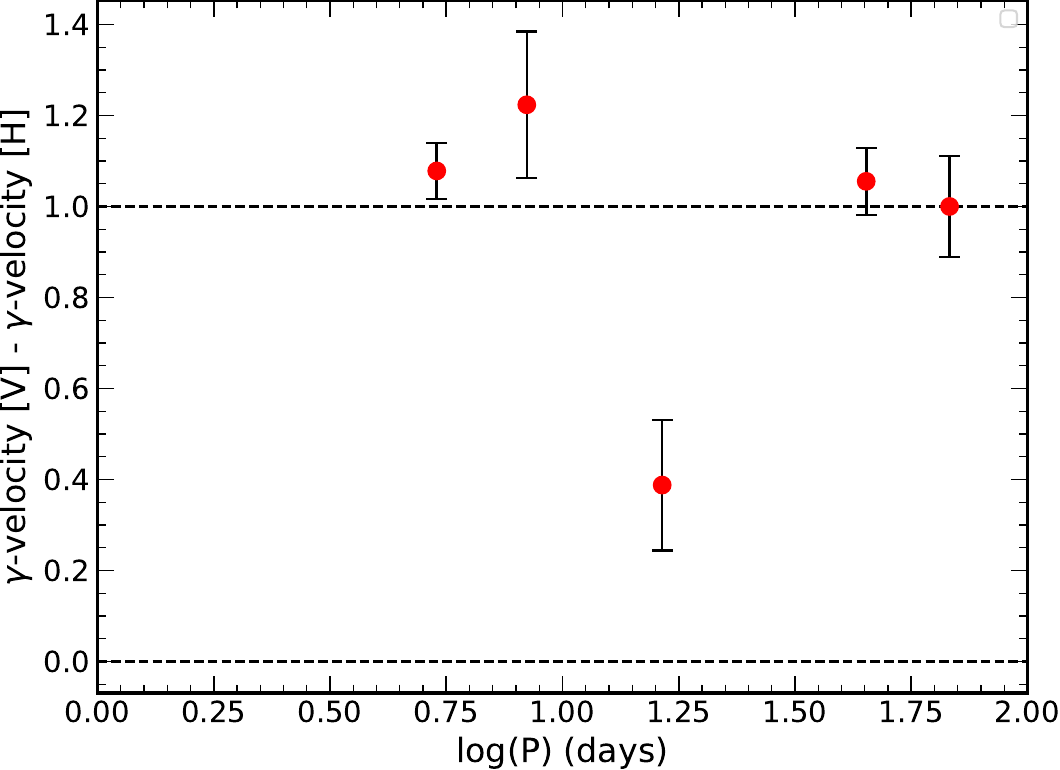}}
\caption{Difference of $\gamma$ velocities between V and H bands for the five Cepheids in our sample.}
\label{fig_Vg}
\end{figure}

In Figs. \ref{fig_fc} and \ref{fig_Vg}, we plot the $f_\mathrm{VH}$ and $\Delta V_{\mathrm{\gamma}}$ quantities as a function of the logarithm of the period, respectively. We obtain a mean value of $f_\mathrm{VH}$ of 1.056 for the five Cepheids with an RMS of 0.016, indicated, respectively, by a horizontal red line and a red area in Fig.~\ref{fig_fc}. It basically means that the RV curve in the V band has an amplitude that is roughly 5\% larger than the RV curve in the H band. In the next section, we discuss a possible physical interpretation of this result. In Fig. \ref{fig_fc}, we add an observational result presented in \citet{nardetto09}, which is the amplitude ratio of cross-correlated radial velocities using a Gaussian fit ($RV_{\mathrm{cc-g}}$) they obtained considering lines located in the middle (around 550nm) and at the edges, respectively, 400 and 700nm, of the V band, as a function of the period of the Cepheid. These results are presented by dashed blue (400 nm with respect to 550nm) and orange (700 nm with respect to 550nm) lines. In this work we compare RV curves in the H band (around 1600nm) and in the visible (typically 550nm). It is interesting to note that our measurements (in the period range of the Cepheids in our sample) are in agreement with the dashed orange line corresponding to a comparison between 700nm and 550nm, despite the slightly different effective wavelengths of this study.

Regarding $\Delta V_{\mathrm{\gamma}}$ we obtain values slightly above 1\kms for all stars except X Cyg for which we obtain a value around 0.4\kms. As indicated in Tab.~\ref{tab_res}, we obtain the highest degeneracy of -0.767 for X Cyg between the two fit parameters, which might explain partly the lower $\Delta V_{\mathrm{\gamma}}$ value of this star.

The fact that H-band cross-correlated RV curves $RV_{\mathrm{cc-g}}$ are about 1\kms blueshifted compared to V-band RV curves could be attributed to a chromatic differential effect of the granulation. Indeed, fundamentally, granulation can blueshift spectral lines because brighter regions (granules) have more material moving toward the observer, while darker intergranular lanes have less material receding toward the star. This effect is seen on the sun \citep{dravins82}. In the more complex case of Cepheids, it has been shown by \citet{nardetto08a} that the granulation affects simultaneously the line asymmetry and the $\gamma$ velocity. This provided for instance an explanation to the so-called k-term of Cepheids, later confirmed by hydrodynamical simulations of \citet{vasilyev17}. A second effect is that this shift can be more or less pronounced depending on where the line is forming within the atmosphere:  
the deeper the line-forming region, the more the mean RV is blueshifted \citep{vasilyev17}, which could indicate that the H-band lines form in the lower part of the atmosphere compared to visual ones.

\begin{table}[htbp]
\begin{center}
\caption{Amplitude ratios and offsets of V and H band RV curves for the Cepheids in our sample.}
\begin{tabular}{lcccccc}
\hline
Star & $f_\mathrm{VH}$ & $\sigma_\mathrm{f_\mathrm{VH}}$ & $\Delta V_\mathrm{\gamma}$ & $\sigma_{\Delta V_\mathrm{\gamma}}$ & $\chi^2_r$ & $\rho$ \\
\hline
$\delta$ Cep & 1.044 & 0.006 & 1.078 & 0.062 & 1.860 & -0.023 \\
S Sge & 1.044 & 0.012 & 1.223 & 0.161 & 8.001 & 0.699 \\
X Cyg & 1.054 & 0.008 & 0.388 & 0.144 & 5.186 & -0.767 \\
SV Vul & 1.051 & 0.005 & 1.055 & 0.073 & 4.050 & 0.164 \\
S Vul & 1.086 & 0.012 & 1.000 & 0.110 & 3.995 & -0.269 \\
\hline
\end{tabular}
\label{tab_res}
\end{center}
\textbf{Notes:} Slope ($f_\mathrm{VH}$) and zero-point ($\Delta V_{\mathrm{\gamma}}$) of the linear relation shown on right panels of Figs.  \ref{fig_RVcc1} and \ref{fig_RVcc2} for the Cepheids in our sample together with their associated uncertainties. The reduced $\chi^2$ is indicated as well as the degenerency ($\rho$) between the two parameters.
\end{table}

\section{The geometric projection factor of Cepheids}\label{s_po}

\begin{figure*}[htbp]
\begin{center}
\resizebox{0.7\hsize}{!}{\includegraphics[clip=true]{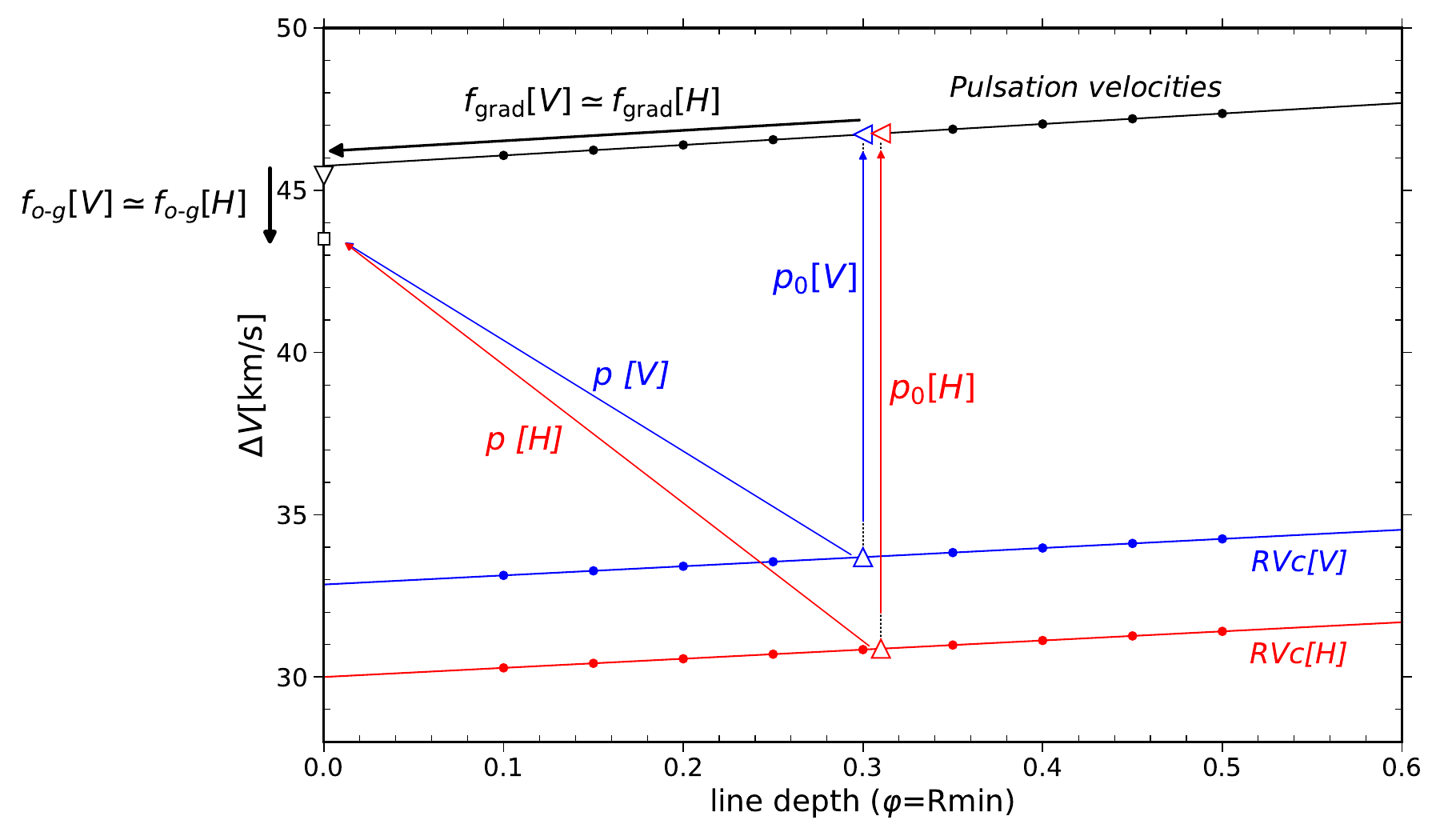}}
\caption{Projection factor decomposition proposed by \citet{nardetto07} applied in V and H bands in a schematic diagram. See the text for explanations.}
\label{fig_decomp}
\end{center}
\end{figure*}

It is well known that the limb darkening combined with the pulsation velocity is the first physical quantity explaining the spectral line shape of Cepheids \citep{vanhoof52}. In this section we use the projection factor decomposition presented in \cite{nardetto07} to analyze and understand the observational results of previous section. \citet{nardetto07} have shown that the projection factor can be decomposed into three quantities (see also \citealt{nardetto23b}): 

\begin{equation} \label{Eq_pf_decomposition}
p= p_{\mathrm{0}}\,f_{\mathrm{grad}}\,f_{\mathrm{o-g}}
,\end{equation}

\noindent where $p_{\mathrm{0}}$ is the geometrical projection factor directly linked to the limb darkening, $f_{\mathrm{grad}}$ is the correction to the projection factor due to the atmospheric velocity gradient, or, said differently, due to the difference of velocity amplitude between the photosheric layer and the layer associated with the line-forming region, and $f_{\mathrm{o-g}}$ a  correction due to the fact that we usually combine interferometry or photometry (sensitive to the optical layer associated with the photosphere) and spectroscopy, mainly sensitive to the motion of the gas into the line-forming region. This is summarized in a schematic diagram in Fig.~\ref{fig_decomp}. In this figure, the blue dots are arbitrary and indicative (for a pedagogical purpose) and correspond to the amplitude of the radial velocities associated with different spectral lines of different line depths (taken at $\phi = R_\mathrm{min}$ to ensure a correct extrapolation to the photopheric velocity; see \citet{nardetto07} for further explanations). The open blue triangle corresponds to the cross-correlated RV (which can be considered in first approximation as a mean of the individual spectral lines). Then, we have 

\begin{equation} \label{Eq_pf_decomposition_V}
p [V] = \frac{V_\mathrm{puls}[V]}{RV_\mathrm{cc-g}[V]} = p_{\mathrm{o}} [V] \,f_{\mathrm{grad}}[V] \,f_{\mathrm{o-g}}[V]
.\end{equation}

We can now add red dots (again arbitrary) that correspond to the lower amplitude RV amplitudes obtained in the H band (note that Y axis values are also arbitrary). Similarly we have

\begin{equation} \label{Eq_pf_decomposition_H}
p [H] = \frac{V_\mathrm{puls}[H]}{RV_\mathrm{cc-g}[H]} = p_{\mathrm{o}} [H] \,f_{\mathrm{grad}}[H] \,f_{\mathrm{o-g}}[H]
.\end{equation}

At this stage, we can make several assumptions. First, we assume that the pulsation velocity associated with the optical layer of the photosphere in the V band is similar to the one in the H band, so that $V_\mathrm{puls}[V] \simeq V_\mathrm{puls}[H]$. This is confirmed at the 1-2\% level from hydrodynamical simulations as tested in \citet{nardetto11b}, but in the case of V versus the K band. Second, we assume that $f_{\mathrm{o-g}}[V] \simeq f_{\mathrm{o-g}}[H]$ is the same, meaning that the optical versus gas layers differential motion in the V or H band is similar. Third, we assume that $f_{\mathrm{grad}}[V] \simeq f_{\mathrm{grad}}[H] $, which is equivalent to consider that the barycentric atmospheric layer associated with the cross-correlated profile (or the mean of all spectral lines considered in the procedure) is located roughly at the same position in the atmosphere of the star in the V and H bands, respectively. This is a reasonable assumption, as pointed out by figures 21 and 22 of \citet{sasselov90}. However, as has already been mentioned, the H-band line-forming regions probably form deeper in the atmosphere because of their blueshifted radial velocities compared to the V band. We believe that the impact of convection and the location of infrared spectral lines in the atmosphere has a stronger impact on the spectral line asymmetry (and $\gamma$ velocity) than on the amplitude of the RV, and thus on the $f_{\mathrm{grad}}$ quantity, which supports the validity of this assumption. Basically, with all these hypotheses, we assume that the impact of the dynamical structure of the atmosphere in V and H onto the projection factor is, at first order, similar, which is also confirmed by additional tests presented at the end of this section. Then, with all these hypotheses, by taking the ratio of Eqs.~\ref{Eq_pf_decomposition_V} and \ref{Eq_pf_decomposition_H}, we obtain 

\begin{equation}\label{Eq_pf_final}
f_\mathrm{VH} = \frac{RV_\mathrm{cc-g}[V]}{RV_\mathrm{cc-g}[H]} = \frac{p [H]}{ p [V]} =  \frac{ p_\mathrm{0}[H]}{ p_\mathrm{0}[V]}
.\end{equation}

\begin{figure}[htbp]
\centering
\includegraphics[width=0.45\textwidth]{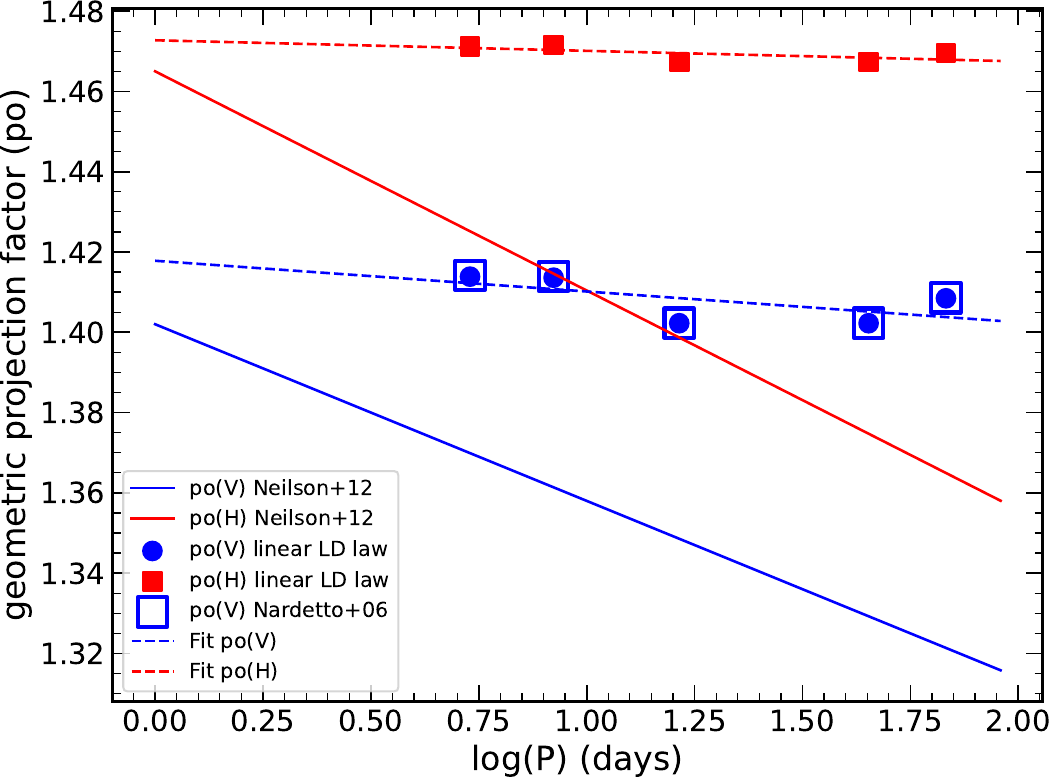}
\caption{Geometric projection factor ($p_\mathrm{0}$) of the Cepheids in our sample as a function of the logarithm of the period. The relation provided by \citet{neilson12} in the V and H band are shown with solid blue and red lines, respectively. The $p_\mathrm{0}$ values for the five Cepheids in our sample were also derived using the linear limb darkening laws of \citet{claret11} and the classical relation from \citet{getting34} (see text for explanations). The ratio of $p_\mathrm{0}$ values in V and H bands presented in this plot are compared to observations in Fig.~\ref{fig_fc}.}
\label{fig_Ppo}
\end{figure}

To test Eq.~\ref{Eq_pf_final}, we calculated $\frac{ p_\mathrm{0}[H]}{ p_\mathrm{0}[V]}$ using the $Pp_\mathrm{0}$ relations of \citet{neilson12} (linear model indicated in their Table 1) based on spherically symmetric SATLAS stellar atmosphere models \citep{lester08}. The $Pp_\mathrm{0}$ are plotted in Fig.~\ref{fig_Ppo} by the solid blue (V band) and red (H band) lines. The ratio of these two curves is plotted in Fig.~\ref{fig_fc} by a dashed red line. This relation is consistent with our observational results. The only exception is S Vul, the longest-period Cepheid in our sample. Such a Cepheid might have strong dynamical effects in its atmosphere, which probably behave differently in the V and H bands. Also cycle-to-cycle variations are not excluded, as is shown by Fig.~\ref{fig_RVcc2}.
 
We tried another approach to verify Eq.~\ref{Eq_pf_final}. We used the $T_\mathrm{eff}$ and $\log g$ values of the star in our sample available in \citet{gro20b}. We considered a solar metallicity for all stars as well as a microturbulence velocity of 1 \kms. We then used the tables of \citet{claret11} in order to obtain the linear limb darkening coefficients of the intensity distribution of the star in the V ($u_\mathrm{V}$) and H band ($u_\mathrm{H}$), respectively. We finally used the analytical equation from \citet{getting34} to link the geometric projection factor to $u_\mathrm{V}$ and $u_\mathrm{H}$. The results for the five Cepheids are shown in Fig.~\ref{fig_Ppo} by blue (V band) and red (H band) squares. The difference obtained with \citet{neilson12} is expected as the atmosphere models used by \citet{claret11} are plane-parallel. As a check of consistency, we overplot with open blue squares the geometric projection factor using Eq.~6 proposed by \citet{nardetto06a} based on simulations. In Fig. \ref{fig_Ppo}, the dashed blue and red lines correspond to linear fits and their ratio is shown again in Fig. \ref{fig_fc} with a dash-dotted red line. Again, an agreement is found between this relation and the mean value of $f_\mathrm{VH}$ (within the RMS).

To observationally validate our working hypothesis, we did another test. We considered the recent results presented in \citet{bailleul26}. In this study, they used CHARA interferometric observations of Cepheids to measure for the first time the limb darkening coefficients in the R (very close to the V band considered in this work) and H bands of three Cepheids, including $\delta$ Cep. They could then retrieve the geometrical projection factors directly from these coefficients: $p_\mathrm{0} [R] \simeq p_\mathrm{0} [V]= 1.420 \pm 0.016 $ and $p_\mathrm{0 } [H] = 1.461 \pm 0.005$. By applying an inverse parallax of pulsation method to $\delta$~Cep -- using the star's distance from \citep{kervella19b},  the angular diameter curve from \citet{bailleul26}, and the RV curves in V and H bands presented here -- we can derive $p [R] = 1.292 \pm 0.064 $ and $p [H] = 1.331 \pm 0.036$. As a consequence, and from the definition of the projection factor decomposition, we have: $\,f_{\mathrm{grad}}[R] \,f_{\mathrm{o-g}}[R] = 0.922 \pm 0.047$ and $\,f_{\mathrm{grad}}[H] \,f_{\mathrm{o-g}}[H] = 0.911 \pm 0.025$. These values are close, which basically means that the dynamical parts of the projection factors in the R and H bands are similar and that the hypothesis presented above is correct, at least for $\delta$ Cep, but most probably also for other Cepheids in our sample (with the only exception perhaps being S Vul, as has already been mentioned). If we assume that the relative dynamical effects in V and H are similar in the atmosphere of the Cepheids in our sample (which seems to be confirmed by our previous test), we can conclude from this section that the cycle-averaged wavelength dependency of the limb darkening of Cepheids as derived from stellar atmosphere models is, at first order, consistent with the 5\% increase in amplitude observed in the cross-correlated velocity from the H band to the V band.

\section{Conclusion}\label{s_conclusion}

We present a comparison of the cross-correlated RV curves of spectroscopic measurements obtained in the V band (HARPS-N) and H band (GIANO-B) simultaneously. We found that for the five stars in our sample, the V-band RV curve has an amplitude about 5\% larger than in the H band. This can be explained at first order by the wavelength dependency of the geometric projection factor, which is itself directly linked to the expected limb darkening in V and H bands. Besides, the H-band RV curves are blueshifted by about 1 \kms (except for X Cyg) compared to the optical. This is attributed to the interplay of the granulation effect on the line asymmetry and $\gamma$ velocity, and the fact that the lines are forming in the lower part of the atmosphere in the infrared. At this stage, we focus on the cross-correlated radial velocities that provide the highest signal-to-noise ratio. In forthcoming studies, we shall investigate in detail the position and shapes of the individual lines in the optical and infrared in order to obtain additional insights into the projection factor of Cepheids. 

As a final note, we emphasize that the difference in amplitude between the V and H bands is not a fundamental limitation of the parallax of pulsation method. Instead, it implies that a RV measurement obtained in a specific band must be paired with its corresponding projection factor. Furthermore, this work demonstrates that for Cepheids, and pulsating stars in general, it is possible to infer the differential stellar limb darkening in two bands by comparing the amplitudes of their respective RV curves. Similarly, the distinct $\gamma$-velocity offsets observed in the V and H bands are typically addressed in the parallax of pulsation method by normalizing the average of the RV curve to zero.

\begin{acknowledgements}
The authors acknowledge the support of the French Agence Nationale de la Recherche (ANR), under grant ANR-23-CE31-0009-01 (Unlock-pfactor) and the financial support from ``Programme National de Physique Stellaire'' (PNPS) of CNRS/INSU, France. Based on observations made with the Italian {\it Telescopio Nazionale Galileo} (TNG) operated by the {\it Fundaci\'on Galileo Galilei} (FGG) of the {\it Istituto Nazionale di Astrofisica} (INAF) at the {\it  Observatorio del Roque de los Muchachos} (La Palma, Canary Islands, Spain). The observations leading to these results have received funding  from the European Commission's Seventh Framework Programme (FP7/2013-2016)  under grant agreement number 312430 (OPTICON). The authors thank the GAPS and TNG observers F.~Borsa, L.~Di Fabrizio, R.~Fares, A.~Fiorenzano, P.~Giacobbe, J.~Maldonado, and G.~Scandariato.  This research has made use of the SIMBAD and VIZIER\footnote{Available at http://cdsweb.u- strasbg.fr/} databases at CDS, Strasbourg (France), and of the electronic bibliography maintained by the NASA/ADS system.  WG gratefully acknowledges financial support for this work from the BASAL Centro de Astrofisica y Tecnologias Afines (CATA) PFB-06/2007, and from the Millenium Institute of Astrophysics (MAS) of the Iniciativa Cientifica Milenio del Ministerio de Economia, Fomento y Turismo de Chile, project IC120009.  WG also acknowledges support from the ANID BASAL project ACE210002.

Support from the Polish National Science Center grant MAESTRO 2012/06/A/ST9/00269 and DIR-WSIB.92.2.2024 grants of the Polish Minstry of Science and Higher Education is also acknowledged. AG acknowledges the support of the Agencia Nacional de Investigaci\'on Cient\'ifica y Desarrollo (ANID) through the FONDECYT Regular grant 1241073. 

B.P. gratefully acknowledges support from the Polish National Science Center grant SONATA BIS 2020/38/E/ST9/00486.  This work has made use of data from the European Space Agency (ESA) mission {\it Gaia}, processed by the {\it Gaia} Data Processing and Analysis Consortium (DPAC). Funding for the DPAC has been provided by national institutions, in particular the institutions participating in the {\it Gaia} Multilateral Agreement. The research leading to these results  has received funding from the European Research Council (ERC) under the European Union's Horizon 2020 research and innovation program (projects CepBin, grant agreement 695099, and UniverScale, grant agreement 951549). We thank the anonymous referee for their insightful comments. 
\end{acknowledgements}
\bibliographystyle{aa}  
\bibliography{bibtex_nn} 

\end{document}